\documentclass[11pt,twoside]{article} 
\usepackage{asp2004}
\usepackage{epsf}
\usepackage{psfig}
\usepackage{lscape} 

\begin{document} 
\title{Surface Magnetic Field Distributions of the White Dwarfs PG\,1015+014 and
HE\,1045$-$0908}
\author{F.~Euchner,$^1$ S.~Jordan,$^2$ K.~Reinsch,$^1$ K.~Beuermann,$^1$ and 
B.\,T.~G{\"a}nsicke$^3$} 
\affil{
    $^1$ Institut f\"ur Astrophysik, 
    Friedrich-Hund-Platz~1, D-37077~G\"ottingen\\
    $^2$ Astronom.\ Rechen-Institut, M\"onchhofstr.~12--14, 
    D-69120~Heidelberg\\
    $^3$ Department of Physics, Univ.\ of Warwick, Coventry CV4 7AL, UK}

\begin{abstract} 
We have applied the method of Zeeman tomography to analyze the surface magnetic 
field structures of the white dwarfs PG\,1015+014 and HE\,1045$-$0908 from 
spin-phase resolved flux and circular polarization spectra obtained with 
FORS1 at the ESO VLT. We find for both objects field topologies that deviate 
significantly from centred dipoles. 
For HE\,1045$-$0908, the frequency distribution of magnetic field strengths is
sharply peaked at 16\,MG for all rotational phases covered by our data but
extends to field strengths at least five times this value.
In the case of PG\,1015+014 there are significant contributions to the 
frequency distribution in the range from 50 to 90\,MG with the maximum near
70\,MG. The detailed shape of the frequency distribution is strongly variable
with respect to the rotational phase.
\end{abstract}

\section{Zeeman Tomography}
The optical spectra of magnetic DA white dwarfs are 
dominated by broad Balmer 
absorption patterns which are characteristic for the magnetic field structure 
in the visible part of the photosphere. Our evolution-strategy based Zeeman 
tomography code allows the determination of the surface magnetic field geometry 
from a set of rotation-phase resolved flux and circular polarization spectra 
by fitting theoretical model spectra from a precomputed database on a grid of 
temperature $T$, field strength $B$, and the angle $\psi$ between the 
field direction and the line of sight. 
The method's ability to reconstruct field geometries 
from synthetic data has been demonstrated by Euchner~et~al.\ (2002). 
For the present study, the magnetic field geometry has been parametrized 
using a truncated multipole expansion up to order $l=3$.

\section{Observations}
Our VLT/FORS1 observations were carried out in May (PG\,1015+014), 
Jun and Dec 1999 (HE\,1045$-$0908) in spectropolarimetric (PMOS) mode. In order 
to eliminate Stokes parameter crosstalk, circular polarization ($V$/$I$) spectra
were computed from two subsequent flux spectra taken with the retarder plate 
rotated by $\pm$\,45$^\circ$. We compensated for wavelength-dependent 
seeing-induced flux 
losses by normalizing the flux spectra to a mean continuum level (Fig.~1).

\begin{figure}[!ht]
\plottwo{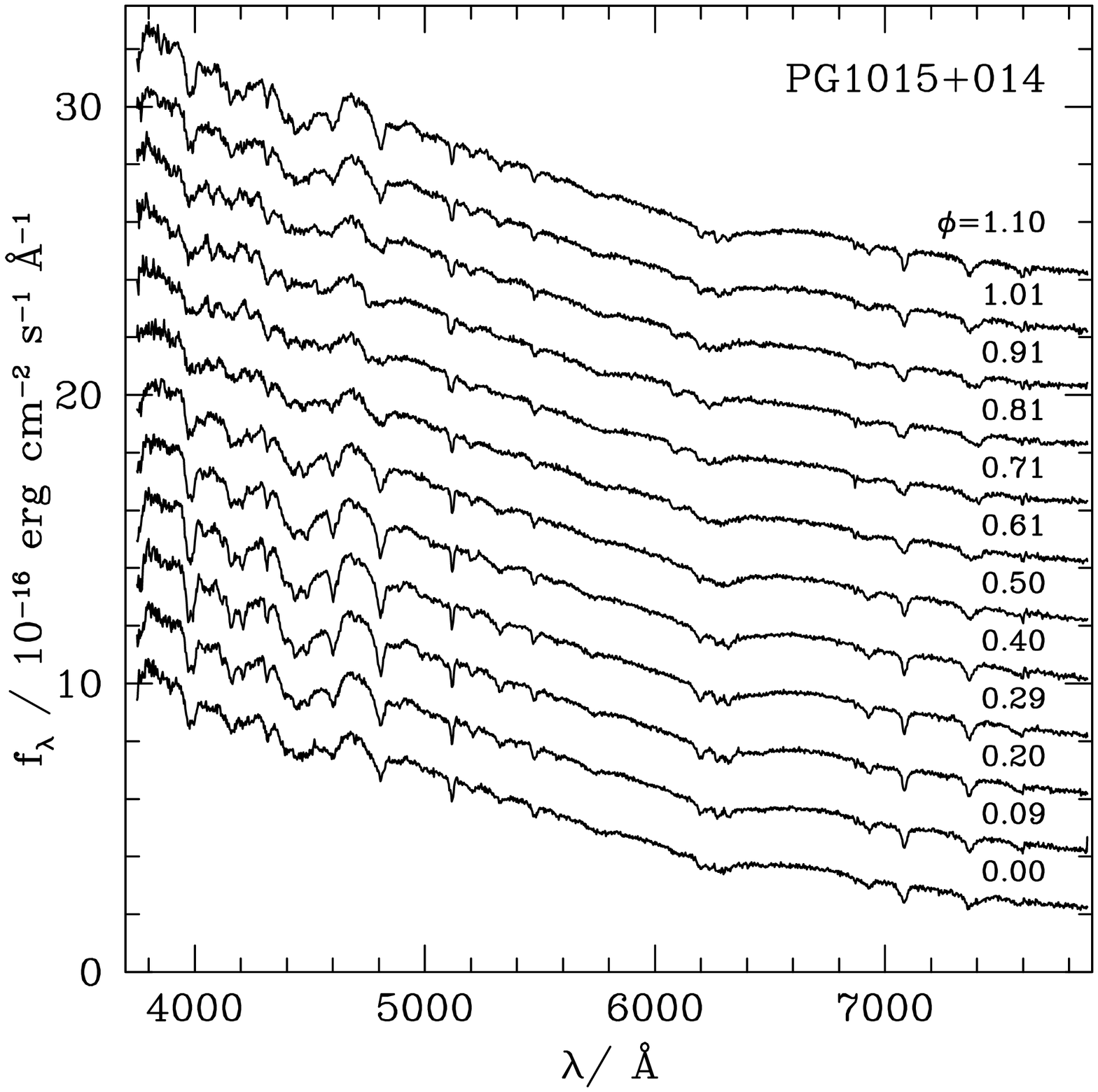}{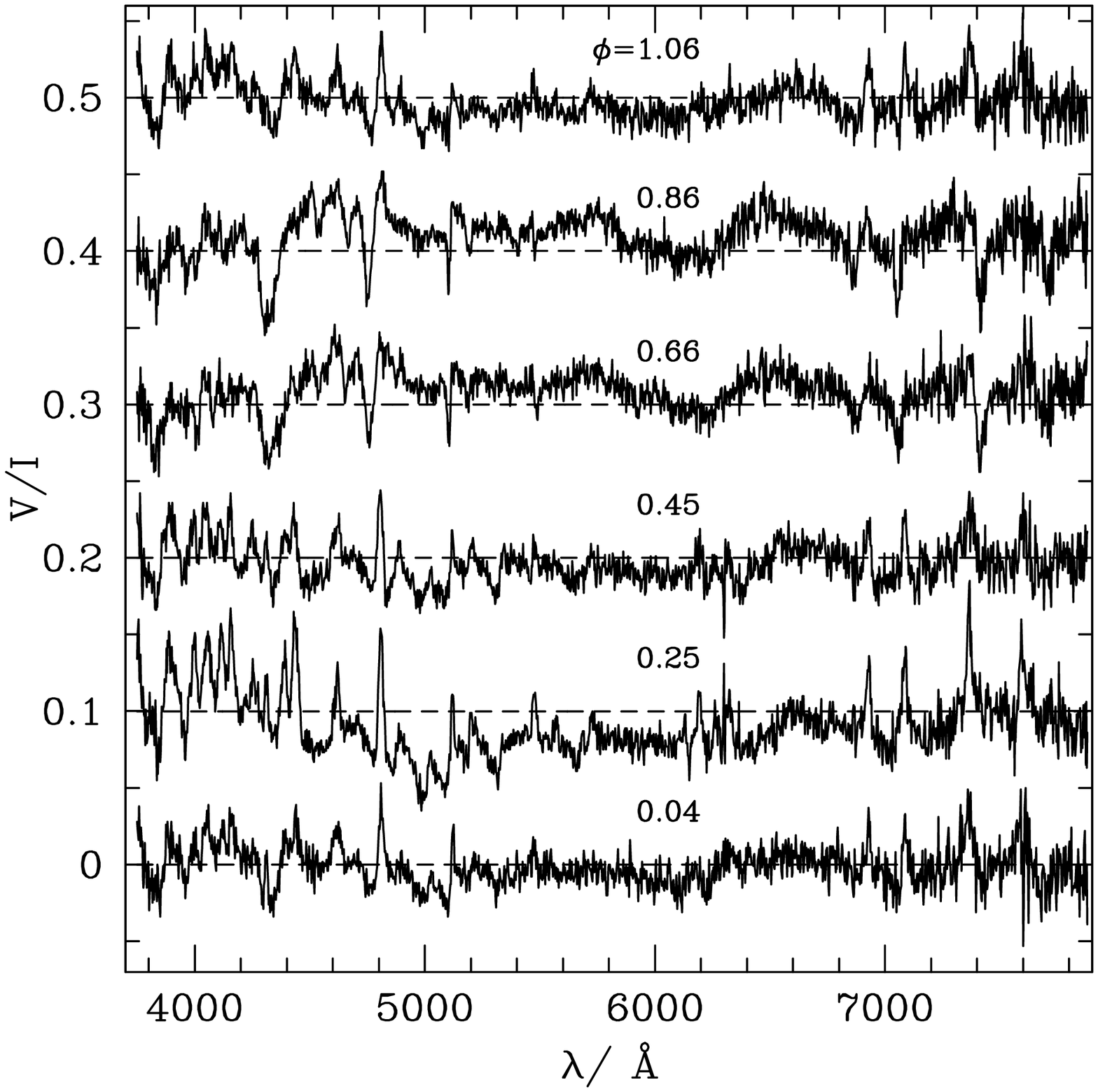}

\vspace*{0.3cm}
\plottwo{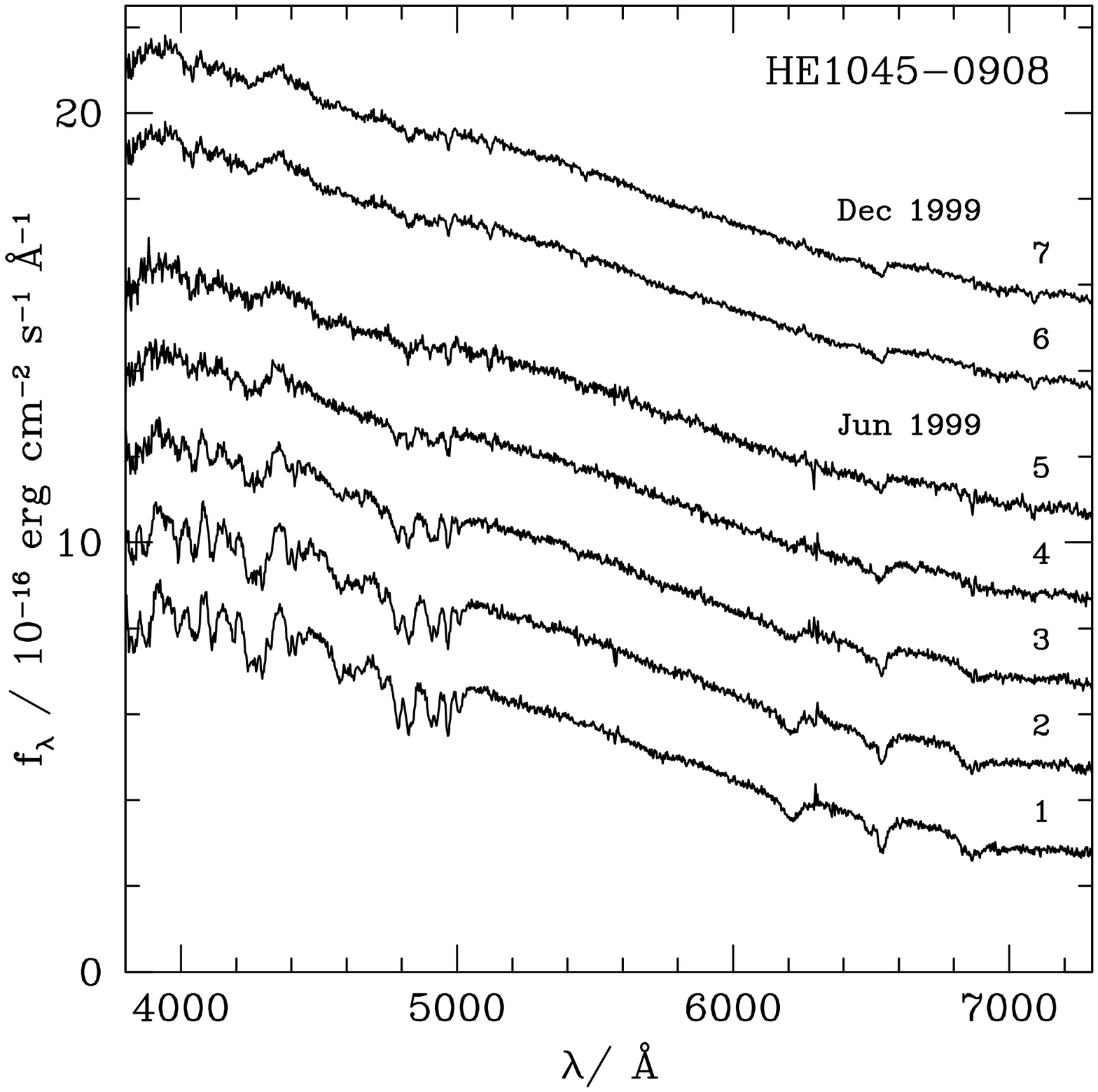}{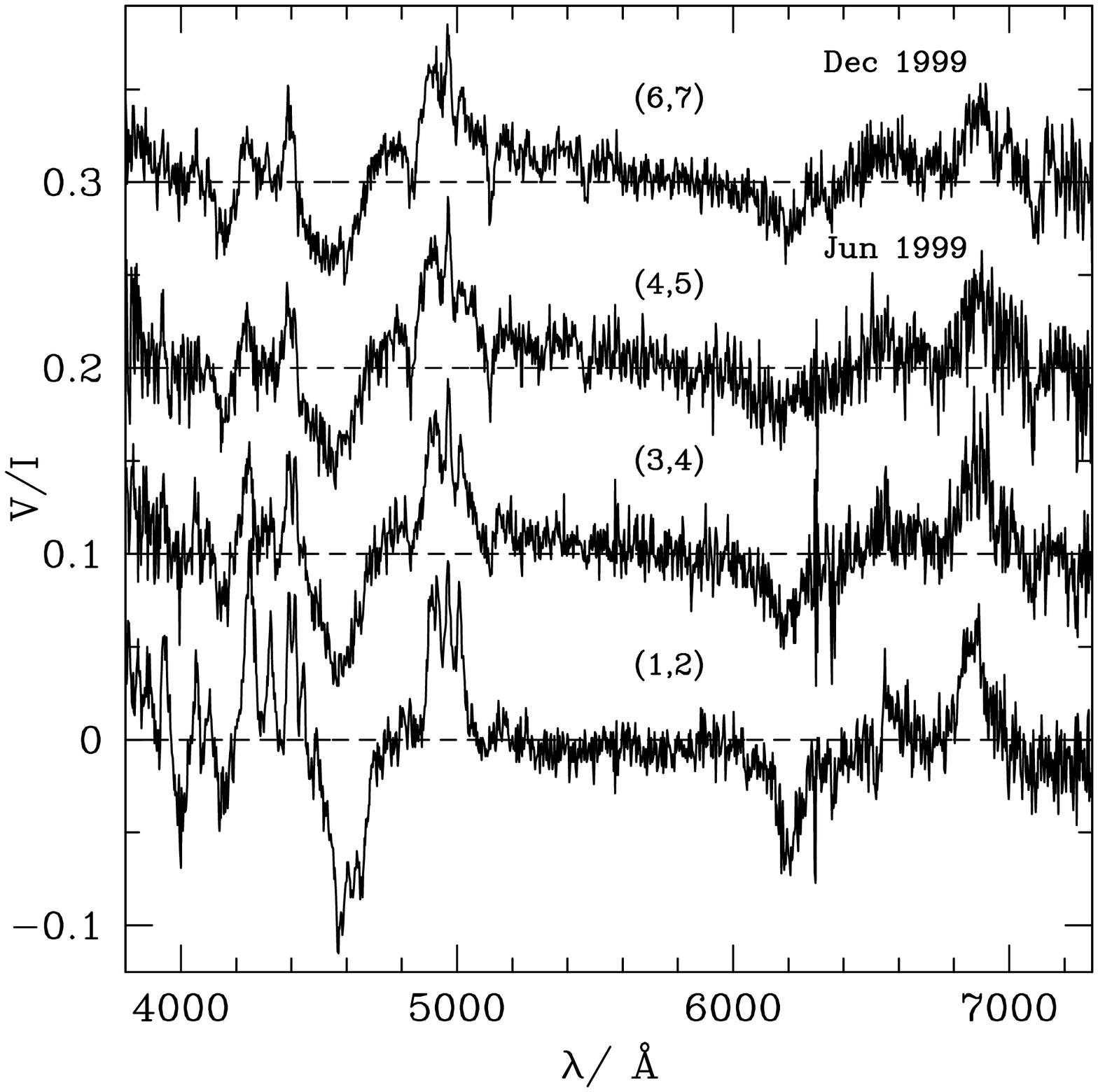}
\caption{Flux and circular polarization spectra of PG\,1015+014 \textit{(top)} 
and HE\,1045$-$0908 \textit{(bottom)} taken with the VLT/FORS1}
\end{figure}

\section{PG\,1015+014}
We collected the spectra into 5 phase bins covering the whole 
rotation cycle according to the ephemeris of Schmidt \& Norsworthy
(1991). The spectra are characterized by sharp, distinct 
Zeeman features without pronounced H$\alpha$ triplet structure, thus indicating 
that the dominating field strengths exceed $B \sim$\,50\,MG.
The best-fit magnetic field structure 
(Fig.~2) 
differs clearly from a simple centred or offset dipole. The frequency 
distribution of field strengths peaks at 70\,MG with an additional peak at 
80\,MG for $\phi=0.66$ and 0.86. Since there are still substantial deviations 
between observed and best-fit spectra, we conclude that the real field 
structure is too complex to be modelled adequately by a truncated multipole 
expansion up to $l=3$. This conclusion seems to be justified because for other 
objects studied by us (e.g. HE 1045$-$0908, see below) the $l=3$ multipole 
expansion yields satisfactory results.

\begin{landscape}
\begin{figure}[!ht]
\plotfiddle{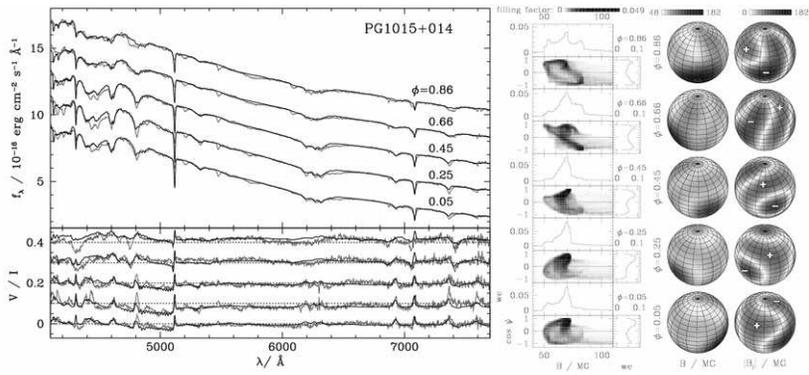}{0.87\textwidth}{0}{36}{36}{-310}{50}
\caption{Field configuration of PG\,1015+014. \textit{Left:} 
Observed (grey line) and best-fit flux and circular
polarization spectra (black line, $T=10$\,kK), 
\textit{centre:} $B$--$\psi$ diagram, 
\textit{right:} surface magnetic field and longitudinal field component}
\end{figure}
\end{landscape} 

\begin{figure}[!ht]
\plotfiddle{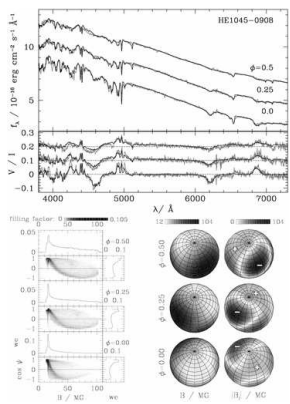}{0.46\textheight}{0}{29}{29}{-105}{0}
\caption{Field configuration of HE\,1045$-$0908. 
\textit{Top:} Observed (grey line) and best-fit flux 
and circular
polarization spectra (black line, $T=9$\,kK), 
\textit{bottom left:} $B$--$\psi$ diagram, 
\textit{bottom right:} surface magnetic field and longitudinal field component}
\end{figure}

\section{HE\,1045$-$0908}
By comparing our observations to those of Schmidt~et~al.\ (2001) it seems 
plausible that our sequence corresponds 
to $\sim$\,0.5 of the rotation cycle.
Thus, we tentatively collect our data into 
3 phase bins (0, 0.25, 0.5). The $\phi=0$ spectrum shows a typical low-field 
H$\alpha$ 
triplet structure generated by a rather uniform field distribution of 
$\sim$\,16\,MG 
that is shallowed out in the subsequent phases by increasing high-field 
contributions (Fig.~3). Although the model fits the overall shape of the 
observed spectra fairly well, some small deviations are still obvious.
An insufficient $B$ resolution of our database in the low field regime 
and our simplified linear limb-darkening law are two possible causes.

\end{document}